# High Resolution Dielectric Characterization of Single Cells and Microparticles Using Integrated Microfluidic Microwave Sensors

Arda Secme[1,2‡], Uzay Tefek[1,2], Burak Sari[4], Hadi Sedaghat Pisheh[1,2], H. Dilara Uslu[1], Ozge Akbulut[3], Berk Kucukoglu[1], R. Tufan Erdogan[1], Hashim Alhmoud[1], Ozgur Sahin[3§], M. Selim Hanay[1,2*]

[1] Department of Mechanical Engineering, Bilkent University, Ankara, 06800, TURKEY

[2] UNAM – Institute of Materials Science and Nanotechnology, Ankara, 06800, TURKEY

[3] Department of Molecular Biology and Genetics, Bilkent University, Ankara, 06800, TURKEY

[4] FENS – Faculty of Engineering and Natural Sciences, Sabanci University, Istanbul, 34956, TURKEY

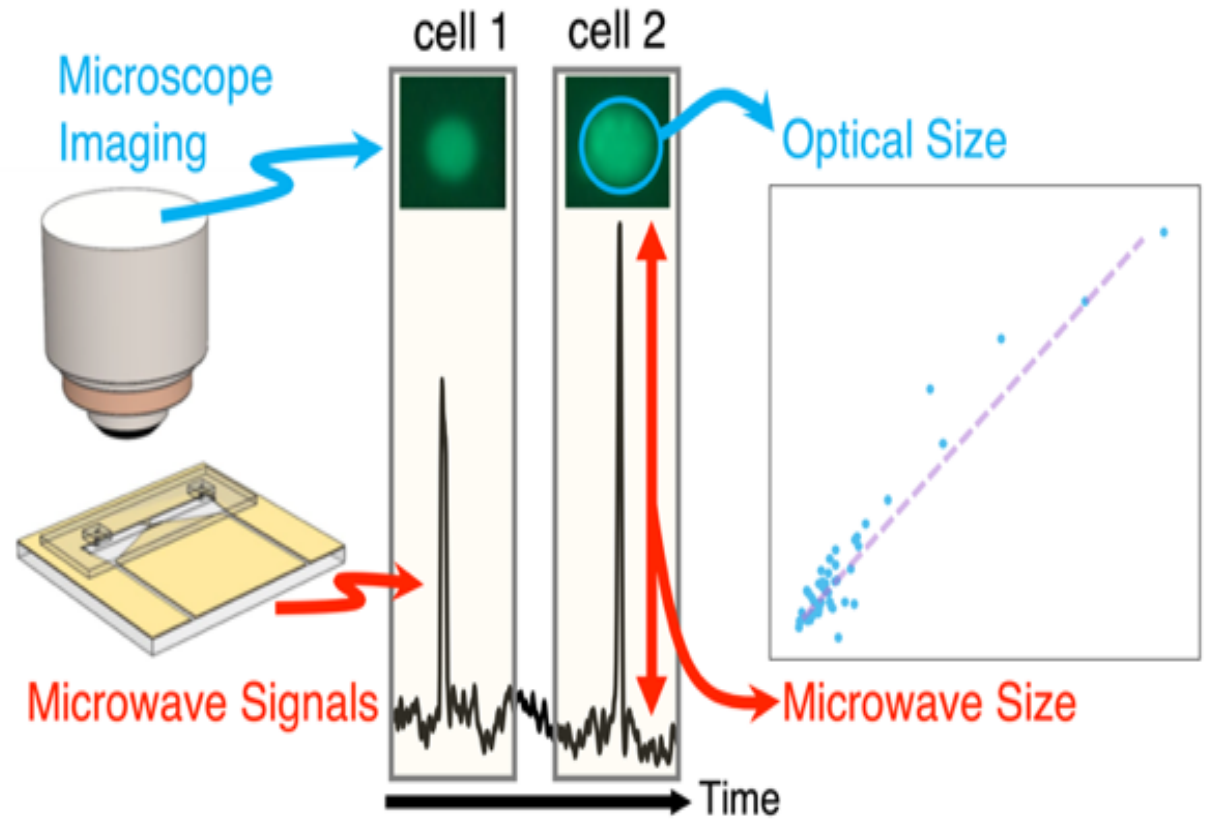

**Abstract:** Microwave sensors can probe intrinsic material properties of analytes in a microfluidic channel at physiologically relevant ion concentrations. While microwave sensors have been used to detect single cells and microparticles in earlier studies, the synergistic use and comparative analysis of microwave sensors with optical microscopy for material classification and size tracking applications have been scarcely investigated so far. Here we combined microwave and optical sensing to differentiate microscale objects based on their dielectric properties. We designed and fabricated two types of planar sensor: a Coplanar Waveguide Resonator (CPW) and a Split-Ring Resonator (SRR). Both sensors possessed sensing electrodes with a narrow gap to detect single cells passing through a microfluidic channel integrated on the same chip. We also show that standalone microwave sensors can track the relative changes in cellular size in real-time. In sensing single 20-micron diameter polystyrene particles, Signal-to-Noise ratio values of approximately 100 for CPW and 70 for SRR sensors were obtained. These findings demonstrate that microwave sensing technology can serve as a complementary technique for single-cell biophysical experiments and microscale pollutant screening.

## 1. INTRODUCTION

Dynamic sizing and material characterization of microscopic particles, such as microplastics and cells, continue to occupy a central role in environmental and biological sciences. For the detection of microplastics in the environment [1, 2], optical microscopy analysis is often used as the first step for feature selection, followed by FT-IR or micro-Raman analysis for material classification. In personalized medicine, precise monitoring of cellular size at high speed has opened avenues for the early diagnosis of diseases [3-6]. While optical microscopes are ubiquitous in laboratories, a system with a smaller form factor, lower cost and ease of use is often needed for point-of-care applications. This need has driven a significant amount of research efforts for developing sensors capable of sizing single cells and particles through measuring either the mass [7-18] or electrical properties, such as impedance change at the high frequency band [19-27] or capacitance at microwave frequencies [28-35].

Critically, the measured *size* in each approach (optical, inertial, capacitive) corresponds to a different physical parameter. With bright-field optical microscopy, the geometrical volume of an object can be extracted (*e.g.*, by estimating the volume through diameter measurements and assuming spherical shape for objects). On the other hand, mass and capacitance sensors provide outputs that depend not only on the geometrical volume but also on the intrinsic parameters of the object as well — mass density and dielectric permittivity, respectively. In principle, such intrinsic parameters can be used for material classification. Especially, the use of electrical permittivity can form the basis of a robust differentiation approach between microplastics *vs.* cells (and other particles of biological origin), since there is a large difference in the relative permittivity values between the two groups: 2-4 for buoyant microplastics *vs.* ~ 40-60 for the internal content of cells [36, 37] (whereby the effect of cell membrane at microwave frequencies is assumed to be small). To obtain permittivity values, one potential avenue is to combine capacitive and optical measurements, so that the volume of an object can be factored out of its capacitance signal.

The electrical capacitance of a cell is related to the total polarizability of the material accumulated inside it. As a cell synthesizes proteins and biological polymers, such as cytoskeletal proteins and phospholipids, the polarizability of the cell is expected to become markedly different than the polarizability of its external environment consisting of water and ions [38-40]. This contrast in polarizability can be measured as a change in the total capacitance of an electrical sensor. Therefore, capacitive measurements can also serve as a standalone platform for tracking relative changes in cellular size, in addition to their complementary use with optical microscopy for simultaneous sizing and material classification.

When capacitive measurements are conducted at low AC frequencies (<1 MHz), extraneous effects such as Debye screening, electrical double layer formation and cell membrane polarization dominate the response from the sample [41]. At higher operation frequencies (e.g. at 50 MHz), the magnitude of the capacitive impedance of cellular membrane drops below the impedance of the cytosol, opening a door for probing internal cellular parameters [42]. However, electrical screening effects due to mobile ions are still at play, and they form electrical double layers at different material interfaces (*e.g.,* between electrodes and liquid, or liquid and cells) [43]. Such interfacial effects due to ions are bypassed when the applied AC frequency is larger than the inverse of the Debye timescale – typically low GHz frequencies. At these frequencies, electrolyte ions cannot follow the electromagnetic field and cease to interfere with the sample's inherent response [28, 44]; however, hydrogen bonds in water can still follow electrical oscillations up to approximately 20 GHz [35]. As a result, a large contrast difference emerges at the lower end of the microwave frequencies (1-20 GHz) between the extracellular medium (mainly composed of water) and the intracellular biostructures that are rich in biopolymers. This contrast difference facilitated the detection of single cells by microwave resonators [28-35]. The ability to discern internal material properties motivates the development of microwave sensors.

In addition to avoiding ion screening effects, sensors working at microwave and Terahertz frequencies offer other benefits such as non-contact measurements through integrated antennas with small form factors [45-49], as well as metamaterial approaches [50-54] where the sensor response can be tuned by engineered subwavelength structures.

Here we demonstrate that the combined use of microwave sensor signals with geometric size measurements from optical microscopy can provide a classification technique for objects with sizes down to tens of micrometers (Figure 1). Analytes with similar geometric sizes, but different electrical permittivity values induce signals of different magnitudes in the microwave sensor. To achieve high resolution in microwave sensing, we used the combination of narrow-band detection and electric field enhancement at the sensing region similar to the approach in [28, 29, 55] but with several differences: a closed-loop system was implemented to track frequency changes, and the entire microwave sensor including the sensing region is fabricated on the same chip. By combining the microwave measurements with geometric size measurements, we showed that mammalian cells and polystyrene microparticles of similar sizes can be differentiated from each other by the combined sensing principle. We also performed continuous electrical size measurements on single cells with microwave sensors to track how the cellular capacitance change after the application of a chemical reagent. Our results demonstrate the emerging potential of microfluidics-integrated microwave sensors (MIMS) for environmental screening and single-cell biophysical measurements.

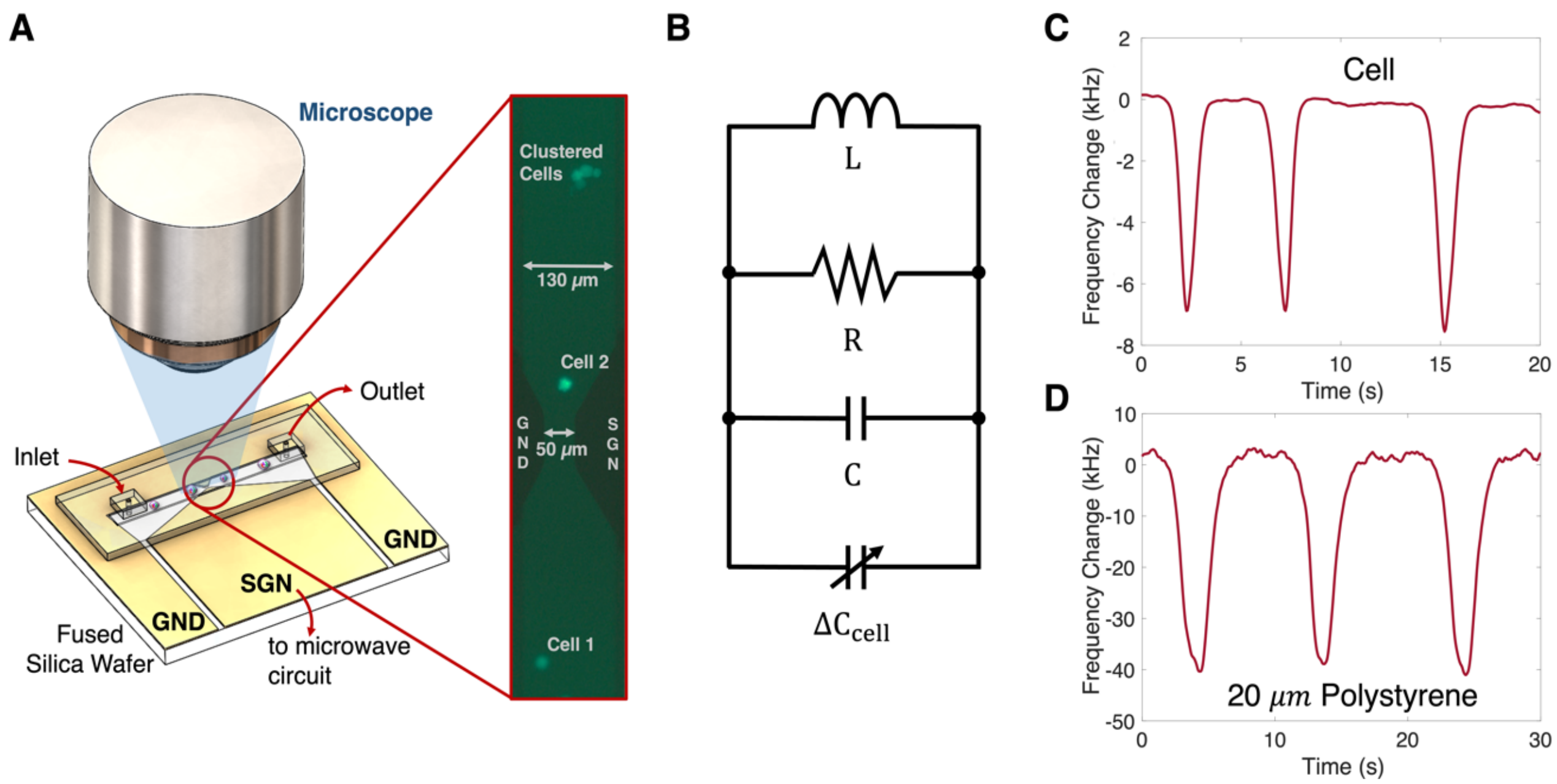

**Figure 1.** (A) Overall schematic of the system: while microwave sensor data was being recorded, the sensing region was simultaneously monitored *via* an optical microscope. A PDMS-based microfluidic channel was bonded to the fused silica wafer and oriented so that the flow passes through the active region of the sensor. (B) The sensor can be modeled as an RLC resonator whereby a passing analyte modifies the capacitance and changes the resonance frequency which is monitored in the experiments. (C) Typical microwave resonator response observed during the transition of single cells (MDA-MB-231-Luc2-GFP). Each precipitous spike is induced by a single cell. (D) Typical response of single polystyrene particles (20 μm nominal diameter) passing through the sensing region.

## 2. EXPERIMENTS

The experimental setup (Figure 1) is composed of several systems: 1) the resonant device for microwave measurements of single cells and microparticles, 2) a custom electronic system for tracking resonance frequency, 3) an optical microscope for independent, optical size measurements, and 4) a fluidic control system that transfers cells to the sensing region

### *A. Resonant Sensor*

A microwave resonator can be modeled as an RLC resonator with an undamped resonance frequency of $\frac{1}{2\pi\sqrt{LC}}$. The resonance frequency can be monitored to convert the RLC resonator into a capacitive sensor: any change in the capacitance of the sensor (*e.g.*, a cell passing in between its electrodes) translates into a change in the resonance frequency of the sensor. Since resonance frequencies can be measured with high precision (e.g. better than $10^{-7}$ as in [56]), the resulting capacitive sensor attains high precision in capacitive detection. For small analytes, the relationship between the induced capacitance change, $\Delta C$, and the resonance frequency change $\Delta f$ can be written, via Taylor expansion, as [40]:

$$\frac{\Delta f}{f_0} = -\frac{1}{2}\frac{\Delta C}{C_0} \qquad (1)$$

where $f_0$ is the original resonance frequency of the sensor, $C_0$ is the original capacitance of the sensor:

$$C_0 = \frac{\epsilon_{eff}}{U_{rms}^2} \int E_{rms}^2(\boldsymbol{r})\, d^3r \qquad (2)$$

where $\epsilon_{eff}$ is the effective permittivity of the resonator, and the integration is performed over the sensor volume. The capacitance change $\Delta C$ can be related to the parameters of the analyte and the sensor as [28, 29]:

$$\Delta C = \epsilon_{fluid}\, 4\pi a^3\, K_{CM} \left(\frac{{E_{rms}}^2(\boldsymbol{r_p})}{{U_{rms}}^2}\right) \qquad (3)$$

Here, $\epsilon_{\mathrm{fluid}}$ is the dielectric permittivity of the host fluid, $a$ is the radius of the analyte, $U_{\mathrm{rms}}$ denotes the root-mean-squared voltage applied to the sensor and $E_{rms}(\boldsymbol{r_p})$ is the electric field at the location of the analyte, $\boldsymbol{r_p}$. Importantly, increasing the local electric field $E_{rms}(\boldsymbol{r_p})$, for instance by decreasing the gap between capacitive electrodes, increases the size of the signal. $K_{CM}$ in equation (3) is the Clausius – Mossotti factor which depends on the permittivity values of the analyte and fluid:

$$K_{CM} = \frac{\epsilon_{analyte} - \epsilon_{fluid}}{\epsilon_{analyte} + 2\,\epsilon_{fluid}} \qquad (4)$$

To demonstrate the versatility of microwave devices, we designed and fabricated two sensors based on two different microwave resonator topologies: **Sensor 1** is a coplanar waveguide (CPW) resonator with a bowtie-shaped sensing region (Figure 2a), while **Sensor 2** is a Split-Ring Resonator (SRR, Figure 2b). Both sensors are fabricated on a fused silica chip and contain active sensing regions where the gap between metal electrodes narrow down (50 μm for sensor 1, and 30 μm for sensor 2) to intensify the local electric field which enhances the detection sensitivity (Figure 2a,b insets). The geometric parameters of the two sensors are detailed on Figure 2. As cells and microparticles pass through the sensing region one by one, they induce distinct frequency shifts in the microwave sensor signal while concurrently being imaged by optical microscopy (Figure 1b, c). The size information derived from optical microscopy then provides the effective radius, $a$, of the particles in equation (2). Thereby, using both the capacitance change $\Delta C$ measured by the microwave resonator, and the effective radius measured optically, one can obtain a quantity proportional to the Clausius-Mossotti factor of the particle.

The coplanar waveguide resonator consists of a lateral gold plane (100 nm thick) patterned on fused silica substrate (250 μm thick). The central conductor forms the signal line whereas the two side conductors form ground lines. These lines are separated by 400 μm throughout most of the device but they narrow down to 50 μm in a bowtie pattern at the sensing region to enhance the local Electric field. The two interfaces of the chip define a resonator structure with a resonance frequency around 2-3 GHz (Figure S2).

The SRR sensor [57, 58] consists of two concentric gold rings as shown in Figure 2b: the outer ring is directly connected to the microwave power source. It inductively couples with the inner ring and excites its resonance modes. Since there is a small gap in the inner ring, the induced current cannot flow through the loop, and for some modes, opposite charges accumulate at the two sides of the gap which produces an intense Electric field used for sensing particles [59]. The dimensions of the resonator are designed to have half-wavelength harmonics in the 4-8 GHz frequency range which is a common spectrum for microwave components (Figure S3). The sensor is patterned on a 500 μm thick fused silica wafer which has low loss tangent and is coated with 150 nm gold.

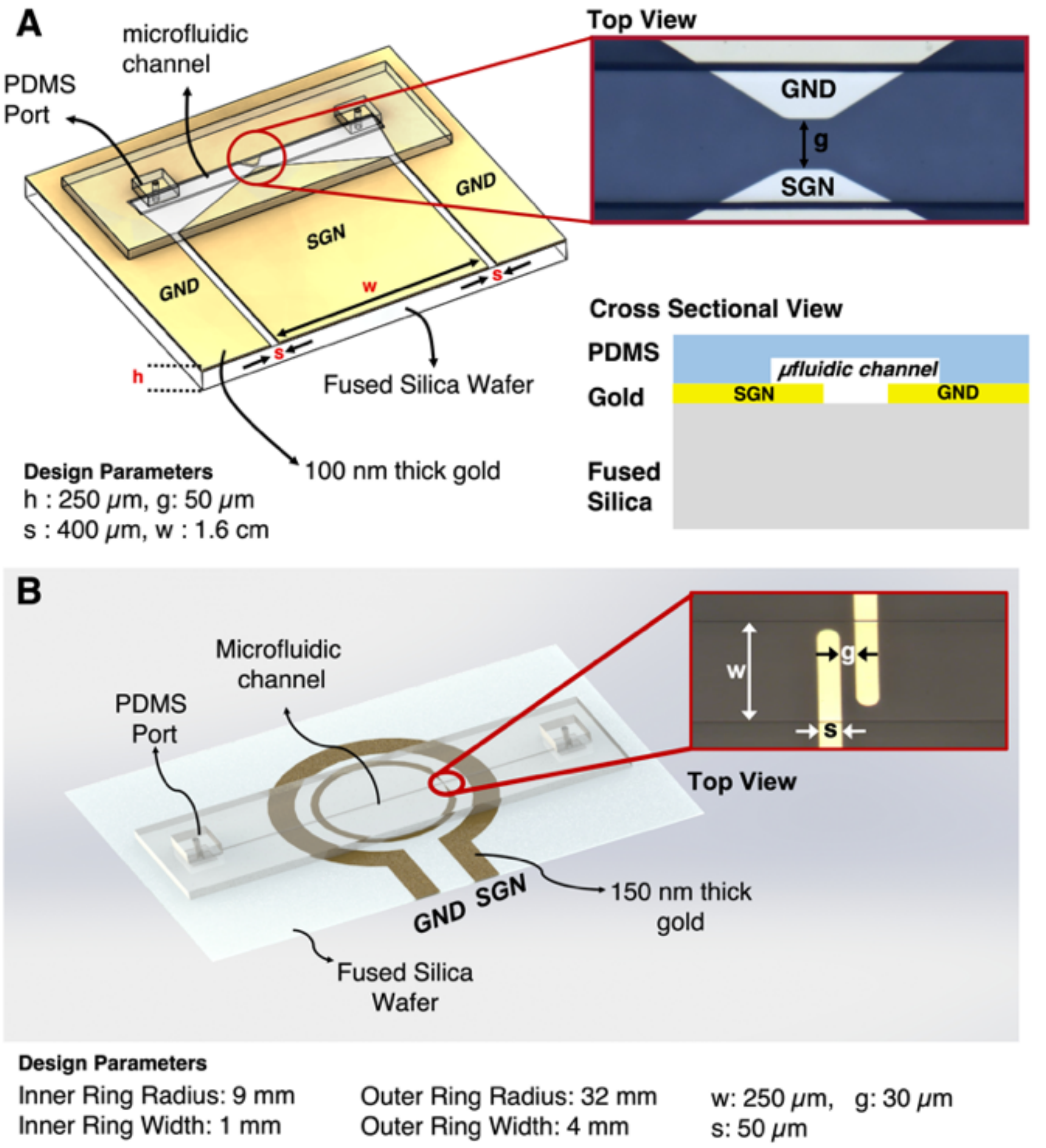

**Figure 2.** Design of (A) CPW, and (B) SRRs sensors. In both cases the signal is fed at the signal port with an SMA connector (not shown in the images). Geometric values are noted down at the figures.

For both type of resonators, a PDMS microfluidic channel is aligned with the capacitive gap and bonded to the silica wafer via an oxygen plasma asher process. CPW sensors with different PDMS channel heights were fabricated and used in the experiments (35, 55 and 110 μm), whereas a single thickness value (55 μm) was used for SRR devices. Even though gold itself is not easily activated to bind with PDMS, the large area of the silica surface and the small thickness (100 nm) of the gold layer facilitates in having leakage-free, conformal coatings. After each fabrication, DI water has been pumped through to check if the flow is stable in the microfluidic channel.

## *B. Measurement Circuitry and Feedback Controller*

The microwave characteristics of each MIMS device (such as the resonance frequency and Quality Factor) were first measured using a vector network analyzer (Figure S2, S3). After this initial characterization, we performed single-particle measurements where we used a custom electronic setup (Figure 3) which contains a signal generator, a detection circuitry and a controller. The controller takes the phase change ($\Delta\phi$) of the resonator as an error signal and uses it to estimate the resonance frequency, which depends on the capacitance change via equation (1). We note that observed phase changes ($\Delta\phi$) are proportional to capacitance changes ($\Delta C$) for a given set of amplitude (R) and phase (ϕ): $\Delta\phi = N(R, \phi)\ \Delta C$ where $N(R, \phi)$ is a normalization factor which stays constant during each experiment since the controller ensures that the same operation point $(R, \phi)$ is tracked.

The detection circuitry (Figure 3) performs narrow-band detection to increase the sensitivity of the MIMS. In this scheme, the narrow-band detection of the resonance frequency was performed with phase-sensitive detection through a lock-in amplifier. Since the maximum operation frequency of the lock-in amplifier (5 MHz) was below the microwave resonance frequency (several GHz), we constructed an external heterodyne circuitry [28, 29] to down-convert the signal to lower frequencies (~ 3 MHz).

In the detection circuit (Figure 3), the driving signal from the signal generator ($\omega_{\mathrm{drive}}$ ~ 2-3 GHz) was first up-converted at Mixer 1 with the reference signal from the lock in amplifier ($\Delta\omega$ ~ 3 MHz). Then, the signal propagates to the circulator to drive and read the response of the resonator. An RF Isolator was placed to reduce the interference due to reflections. After the isolator, the response was down-converted at Mixer 2 and filtered with a low pass filter (4.5 MHz). The output of this detection circuit block, $\phi_{out}$, is the phase difference between the instantaneous phase and the reference phase at the resonance frequency of the device: this difference serves as an error signal for the controller circuit.

The error signal generated is fed to a Proportional-Integrative (PI) controller. The PI controller generates a control signal ($\omega_{correction}$) and sends it to the signal generator via LabView interface. This control signal updates the drive frequency of the microwave signal generator ($\omega_{drive}$) until it matches the resonance frequency ($\omega_{\mathrm{res}}$) of the sensor. In other words, the PI controller sends a command to the signal generator to update its driving frequency by an amount:

$$\omega_{correction}[n] = K_P \phi_e[n] + K_I \sum_{m=0}^{m=n-1} \phi_e[m] \qquad (5)$$

where $n$ and $m$ denote time indices, $K_P$ is the proportional feedback coefficient and $K_I$ is the integral feedback coefficient. At the beginning of the experiments, these coefficients are tuned to obtain desired measurement bandwidth, which was typically 1 Hz in this study.

With the PI controller, the phase angle of the resonator response function was locked at 0 degrees (i.e. the phase value at resonance) to form a Frequency-Locked Loop (FLL). Any deviation of phase away from 0 degrees emerged as an error signal which was used to update the frequency of the signal generator. With this method, we can continuously track the resonance changes with high precision[60].

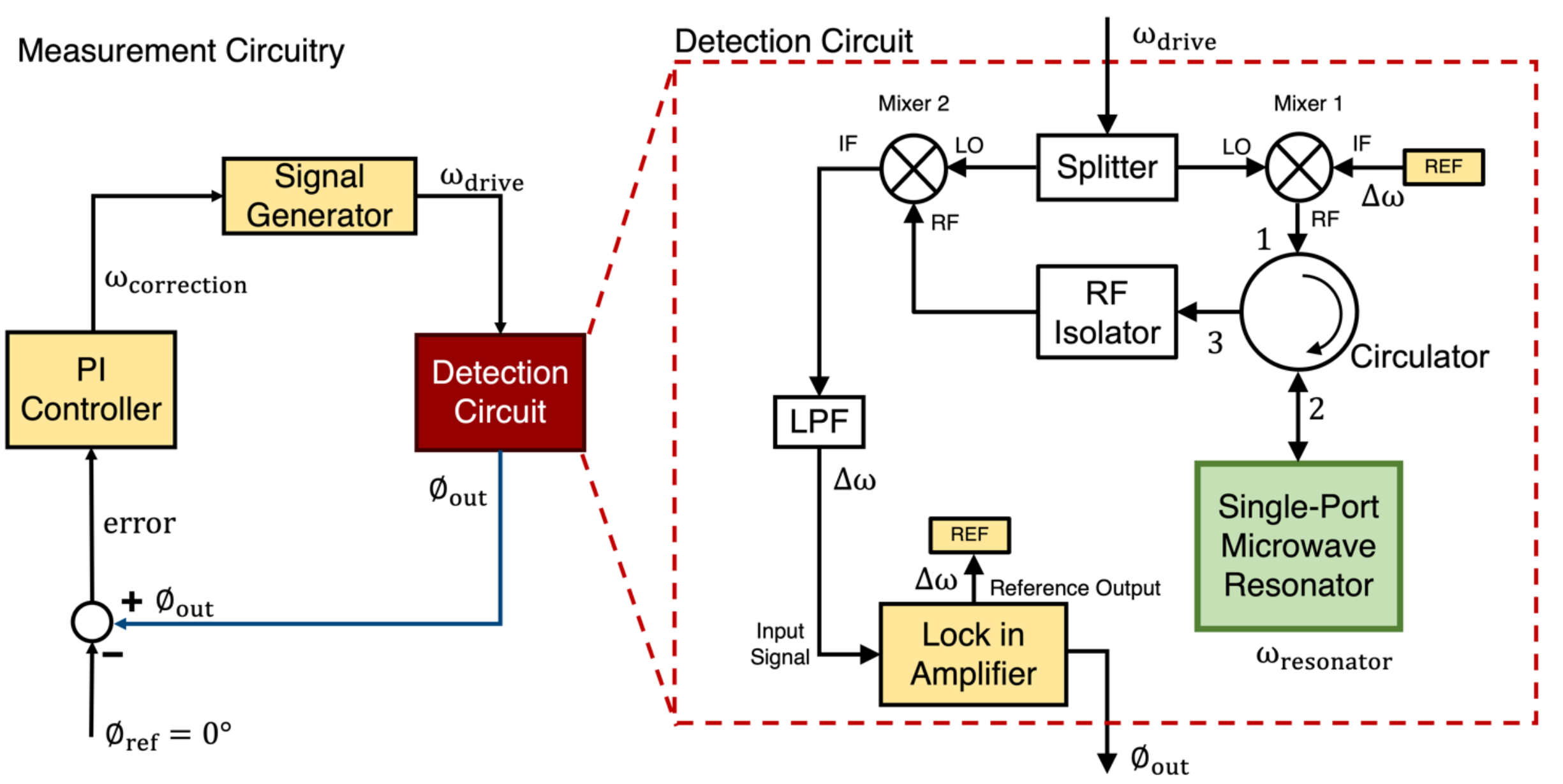

**Figure 3.** Measurement circuit consists of a signal generator to drive the resonator, a detection circuitry to determine the phase of the resonator at the applied frequency, and a PI controller to update the resonance frequency of the signal generator. LPF: Low pass Filter.

## C. *Experimental Setup*

The fabricated MIMS chip, together with microwave and microfluidic components, was placed on the sample stage of a fluorescence microscope. While the sensing region was observed by optical microscopy, mammalian cells (MDA-MB-231-Luc2-GFP) and polystyrene microparticles of varying size (10 μm or 20 μm diameter) were transported through the active region in separate runs (Methods). To transport the analytes, we used a controllable pressure pump (Fluigent; MFCS-EZ) with typical flow rates of $0.3 - 1\ \mu L/min$. These flow rates were chosen so that accurate optical images of cells can be obtained as they passed through the sensing region.

## 3. EXPERIMENTAL RESULTS

### *D. Particle and Cell Characterization Experiment*

For sensor 1, the resonance frequency (at 2.7 GHz) was monitored with an Allan deviation of $4 \times 10^{-8}$ and frequency modulations induced by single cells crossing the sensing region were recorded. While microwave data were being measured, a video of the cells transiting through the sensing region was recorded simultaneously to correlate the optical and microwave signals (Supplementary Video 1). We calculated the geometric volume of the cells and polystyrene microparticles from microscopy images, assuming a spherical shape. We then plotted the geometrical size *vs*. the corresponding microwave frequency shifts to summarize the data in both dimensions (Figure 4A).

As expected, the optical size of the mammalian cell sample spanned a large range: microwave sensor response followed the same trend in an almost linear manner. The data for polystyrene particles were concentrated around the nominal value of the sample (20 µm). Importantly, the data clusters for polystyrene and cells were well separated from each other. Indeed, a polystyrene particle generates several times larger frequency shift in the microwave sensor compared to the frequency shift induced by a cell of similar size.

To demonstrate the general applicability of material differentiation, we conducted a similar experiment with sensor 2 (the SRR device) at 4.2 GHz with an Allan Deviation of $7.5 \times 10^{-7}$. As before, different analytes populated distinct regions in the analytical space formed by optical volume and microwave resonance shift (Figure 4B). As expected, the polystyrene particles generated much larger microwave signals compared to cells of the same optical size.

The microwave signal induced by each analyte depends on how much its permittivity differs compared to the permittivity of the culture media ($\varepsilon_r \sim 78$). More specifically, the signal is proportional to the particle volume and Clausius-Mossotti factor: [17]

$$|\Delta f| = G \times V_{particle} \times K_{CM}(particle) \qquad (6)$$

Here, *G* represents the responsivity of the microwave sensor This can be calculated by combining equations (1), (2), and (3):

$$G = \frac{3}{2} f_0 \frac{\epsilon_{fluid}}{\epsilon_{eff}} \left( \frac{E_{rms}(\boldsymbol{r_p})^2}{\int E_{rms}^2(\boldsymbol{r})\, d^3 r} \right) \qquad (7)$$

An important aspect of the responsivity, G, is the last term inside the parenthesis which is proportional to the local Electric field at the particle location squared over the square of the average electric field throughout the resonator volume. By making the sensing region narrow, this ratio increases. For instance, in the CPW design, the gap between the signal and ground lines is 400 μm on the general CPW structure but this gap decreases to 50 μm in the sensing region.

The estimated responsivity that relates the induced frequency change to the particle volume and Clausius-Mossotti factor is illustrated with dashed lines in Figure 4A with relative permittivity of polystyrene taken as 3 [61] and cells as 50. Clausius-Mossotti factor of polystyrene ($|K_{CM}|$~0.5) is larger than that of cells ($|K_{CM}|$~0.1 at microwave frequencies). Hence, we expect a much larger microwave signal amplitude for polystyrene particles compared to cells with similar size. The overall trend clearly indicates that differentiation of analytes with different compositions is possible by the combined sensing technique.

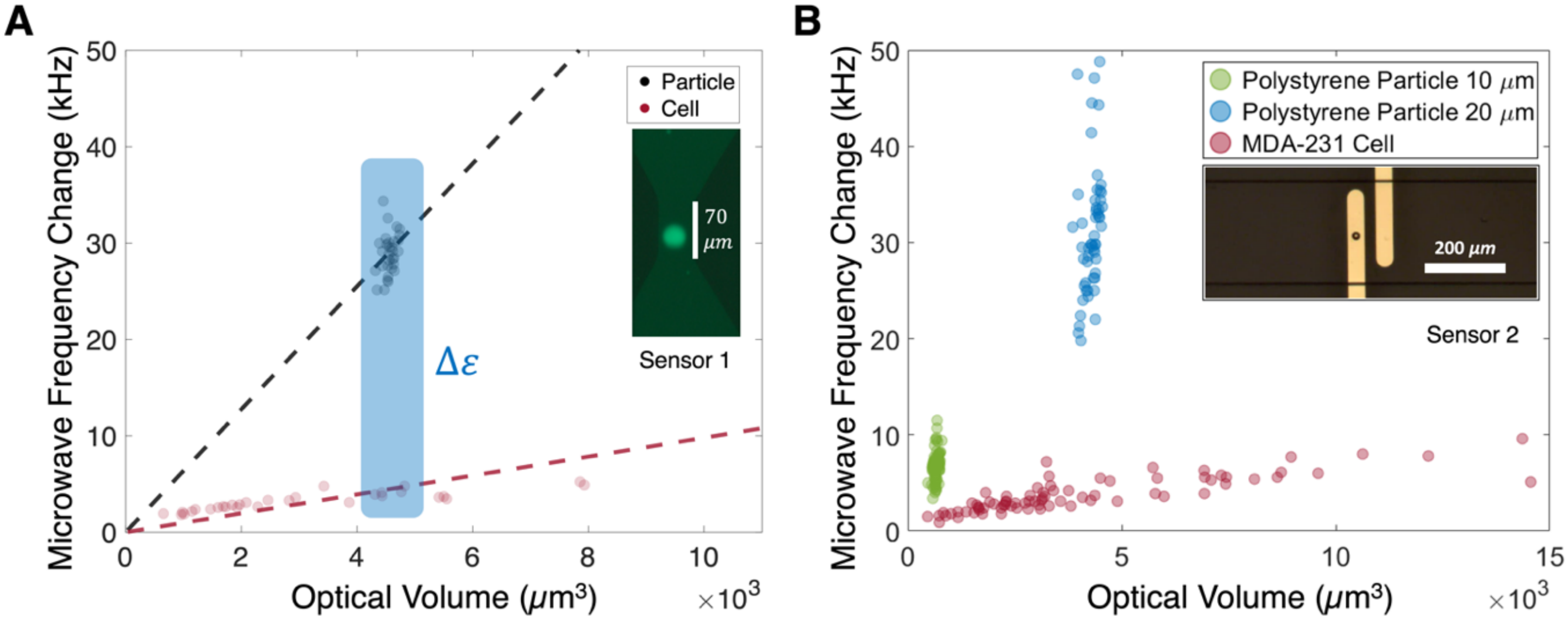

**Figure 4.** (A) Experimental and analytical results obtained by CPW resonator. For cells, we observe a linear relation between optical volume of the cell and microwave frequency change. Large contrast in the dielectric constant can be observed between polystyrene particles and cells having a similar optical volume (blue highlighted region) (B) Same experiment was performed using SRR design with different particles. The same linear relation for the cells can be observed and each type of analyte spans a different region on the analytic plane of optical volume – microwave frequency change. In all cases, the marker sizes are larger than the uncertainties in the single cell/particle measurements. Insets: (A) the active

region of Sensor 1 with a cell under fluorescent microscopy; (B) the active region of Sensor 2 with a 20 μm polystyrene particle.

We attribute the dominant factor behind the dispersion in the frequency shift data to be the vertical-position dependency of the microwave sensor response, as the Electric field distribution is non-uniform (Figure S1) [62, 63]. To support this claim, we repeated the CPW experiments with different microfluidics channel heights: in addition to 55 μm for which the Fig. 2A data were taken, we also used shallower (35 μm) and deeper (110 μm) channel heights (Table I). In each case, the single event resolution (the signal induced by the particle over the noise level) was close to 100. However, when comparing particles on the same ensemble, the dispersion between the measurements were much larger, and ensemble resolutions much lower, as reflected in the ensemble resolution of the column of Table I. We note that with the vendor-specified coefficient of variation for these analytes, the maximum ensemble resolution possible would be approximately 22 (calculations are detailed in the Methods section, SNR Calculations). As can be seen from Table I, ensemble resolution gets better for shallower channels since the particles have less physical space along the vertical direction to spread out when a shallower channel is used.

Resolution level of the SRR sensor (with 55 μm channel height) was similar: for 20 μm particles, mean frequency shift was 31.2 kHz (for a 4.33 GHz resonance) and signal noise was 0.43 kHz , yielding a single particle resolution of 72.

TABLE I

RESOLUTION OF THE CPW RESONATOR WITH 20 μM POLYSTYRENE MICROPARTICLES FOR DIFFERENT CHANNEL HEIGHTS

| Channel Height [μm] | Mean Frequency Shift [kHz] | Noise Level [kHz] | Single Event Resolution | Ensemble Standard Deviation [kHz] | Ensemble Resolution |
|---|---|---|---|---|---|
| 35 | 40.4 | 0.325 | 124 | 2.26 | 17.9 |
| 55 | 28.9 | 0.291 | 99.2 | 2.13 | 13.6 |
| 110 | 12.8 | 0.096 | 134 | 1.18 | 10.8 |

Calculations are detailed in the Methods section, SNR Calculations.

*E. Single-Cell Trapping Experiments*

After the simultaneous use of optical microscopy and microwave sensing, we decided to benchmark the size quantification performance of our microwave sensor. A single cell can be trapped and made to pass back and forth between the sensing electrodes by fine-tuning the applied microfluidic pump pressure dynamically (Supplementary Video 2). The fluorescent optical images of the same cell were recorded during the experiments and then processed by a custom image processing algorithm (Table S1). The algorithm counts the number of bright pixels for 20 consecutive frames and estimates the mean and standard deviation in the geometrical size of the cell. This way, the transition position, dwelling time and velocity of the cell can be obtained while recording its microwave response in real-time (Figure S5).

By recording data for 25 minutes and continuously reversing the applied fluidic pressure, we found that the mean value for the frequency shift was 4.87 kHz (for a resonance frequency of 2.51 GHz) with a normalized standard deviation of 0.09. Here normalized standard deviation, $\hat{\sigma}_{MW}$ , is defined as the standard deviation over the mean (Figure 5). The data does not indicate any directional asymmetry of the sensor for the cell moving forward or backward. Table II summarizes these measurements.

TABLE II

MICROWAVE VS. OPTICAL SIZE MEAUREMENT

| MODE | MEAN | $\sigma$ | $\hat{\sigma} = \frac{\sigma}{MEAN}$ |
|---|---|---|---|
| MICRO-WAVE | 4.87 kHz | 0.44 kHz | 0.09 |
| *OPTICAL* | $1.18 \times 10^4$ μm$^3$ | $0.2 \times 10^4$ μm$^3$ | 0.17 |

The geometric size of the trapped cell was estimated by the image processing algorithm and compared with the frequency shift statistics. We expected the algorithm to yield similar volume values as the same cell was processed in different frames. In fact, in Figure 5, the volume of the cell is concentrated around $1.18 \times 10^4\ \mu m^3$ with a standard deviation normalized to the mean value ($\hat{\sigma}_{opt}$ ) of 0.17, which can be seen at the bottom histogram. In this case where the cell is constantly in motion, the microwave sensor results indicate a relatively narrower distribution for the size of the cell ($\hat{\sigma}_{MW} = 0.09$ for microwave *versus* $\hat{\sigma}_{opt} = 0.17$ for optical measurements, Table II).

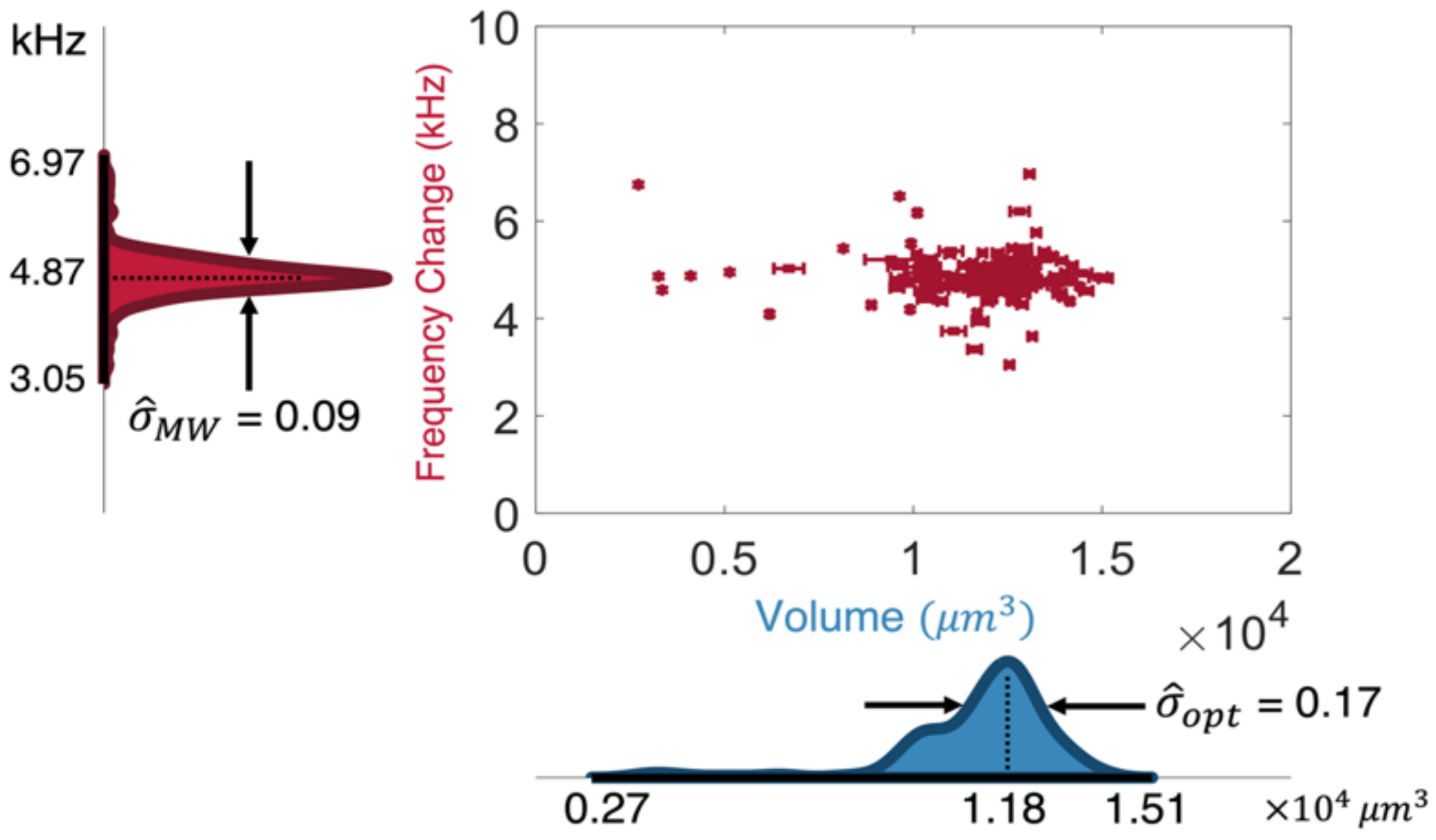

**Figure 5.** Comparison of the size distributions obtained from microwave frequency shift measurements (y-axis) and through fluorescence microscopy (x-axis). There are 171 individual passes from the sensing region, the histograms for volume and frequency shifts are also projected to their respective axes, and in both cases a distinct peak appears as expected.

### *F. Frequency Tracking After Cellular Damage*

The ability to dynamically monitor the effects of drugs and chemicals delivered to diseased cells is of utmost importance. The resulting dynamical effects include changes in volume and dielectric properties of cells. Both factors can affect the frequency shifts measured by the microwave sensors. To this end, we designed another experiment to track the microwave frequency shifts caused by a single cell just after the cell undergoes a sudden morphological change induced by a chemical. As shown in the previous section (Figure 5), the uncertainty in a frequency-based size tracking was found to be approximately 9% of the mean size of the cell, so changes larger than this value can be monitored.

To demonstrate that microwave sensors can track the real-time dynamics of individual cells, we chose to use dimethyl sulfoxide (DMSO), an osmolyte chemical that has been shown to affect cell morphology [64]. DMSO is also known to have dose-dependent toxic effects on cells which scale with the concentration of DMSO, so we expected to see those effects on cells represented in the changes in the resonance frequency shift. We observed that the effect of DMSO is maximized when treating the cells for approximately half an hour in their culture plate with a 50% concentration of DMSO. For the experiment, cells were injected in the presence of DMSO into the sensor microchannel. The DMSO was expected to cause a contraction of the interfacial region between water-lipid boundary over the monitoring period [65]. We expected to see a shrinkage in the lipid bilayer of the cells causing observable shifts in the resonance frequency of the sensor. Indeed, the ten-points moving average of the frequency shifts decreased from 6.43 kHz to 4.64 kHz (Figure 6A) within thirty minutes (28% decrease, well beyond the uncertainty in size measurements). For the control run, the same cell line in another culture dish with the same passage number was processed in the same way, replacing DMSO with culture medium. Using the same device and experimental conditions, another cell from the control population was tracked for more than twenty minutes. Throughout the experiment, the average frequency shift remained close to the original signal level as depicted in Figure 6B (5.46 kHz to 5.11 kHz, a decrease of %7 which is within the measurement uncertainty level).

Even though changes were clearly observed in the frequency shift response of the cells treated with DMSO (relative to untreated cells), it was not clear to what extent those changes were caused by the alterations in the geometrical parameters of cells, or compositional changes in the cell structure. In order to elucidate this matter, we performed verification measurements using flow cytometry on cell samples treated with DMSO as outlined above. First, we verified that after a total of 30 min exposure to DMSO, the cells showed a decrease in viability (84.0%) compared to negative control cells (95.7%) as shown by Propidium Iodide (PI) assay. The increase in the side-scatter (SSC) values (which is a measure of cellular granularity) was most prominent for DMSO treated cells compared to the negative control (Figure S9), while the change in front-scattering (FSC) which is correlated to geometric size was small between the two samples. The change in the microwave frequency shifts being measured seemed to track SSC since the histogram exhibited the largest change between the two samples. Since internal electrical permittivity contributes to the microwave shifts, the measured change in frequency might have been linked to internal changes in the cell due to its exposure to chemical injury (by DMSO) captured as an increase in SSC measurements. In order to verify that visually, cells treated with DMSO were observed under optical microscopy, and compared to untreated cells, where the visual cell diameters were measured for each sample (Figure 6C-E). It was clear that both cell samples (DMSO treated and untreated) had statistically identical diameters ($13.6 \pm 3.8$ μm and $13.8 \pm 3.8$ μm respectively) which agrees well with the FSC results. These measurements were conducted for 150 cells in each sample individually. What was noteworthy was that visually, the DMSO-treated cells appeared to have a disturbed structure with black spots appearing to be located either on the cell membrane itself or internally within the cell cytosol (inset of Figure 6C). This contrasted with the clear translucent interiors of the untreated cells (Figure 6D). Additionally, the DMSO-treated cells appeared non-spherical and exhibited irregular shapes as compared to the mostly spherical geometries of their untreated equivalents. These observations are in agreement with [66] where DMSO treatment substantially altered the morphology of embryoid bodies at concentrations as low as 1% v/v% in a dose-dependent manner, in addition to alterations in attachment behavior and in gene expression.

Even though the diameters of cells were largely equivalent between the two samples statistically, the structural alterations in cell geometry were clearly apparent both through visual inspection and through flow cytometry measurements. This led us to conclude that the information conveyed by the shift in resonance frequency did not only correspond to changes in cell volume but was even more indicative of structural changes in the cell itself. This constitutes a major advantage over other electronic sensing techniques that are only sensitive to clear changes in cell size and the volume it occupies. Microwave sensing in this regard could be seen as analogous to the side-scattering measurements conducted traditionally using flow cytometry and can provide similar information in an electronic fashion rather than optically.

## 4. DISCUSSION

While the microwave sensing approach has been demonstrated in different scenarios, there are several dimensions along which the technique can be further developed. The simultaneous use of optical microscopy with microwave sensors increases the cost of the entire system and analysis duration offered by all-electronic systems; although low-cost [67] and high-throughput [68] optical imaging systems are continually being developed. Indeed, the relatively low flow rates used in this work ($0.3 - 1$ μL/min) are determined by the speed of the standard optical microscopy approach used. On the other hand, the sensing speed of microwave resonator in this case is limited by the communication speed between the function generators and the computer which implements the PI feedback to track the resonance frequency. As a result, typical durations for step response are several hundred milliseconds. Since at this measurement bandwidth, the background drift in the microwave resonance frequency was not significant compared to noise levels, no attempt has been made to decrease the frequency drift of the microwave resonator (e.g. by thermal feedback).

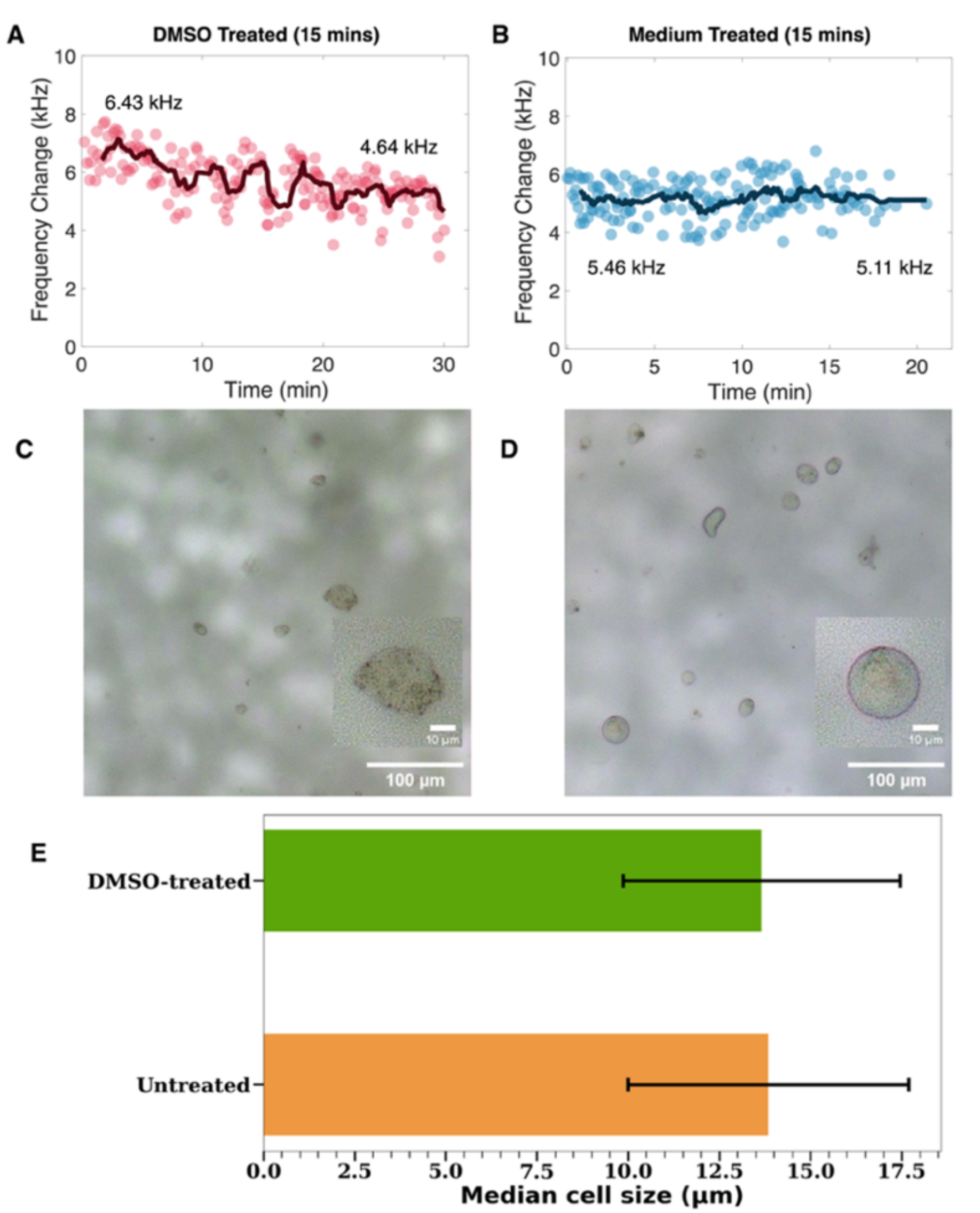

**Figure 6.** Repeated measurements on single cells at the sensing region and the resulting trends in the frequency shifts depending on whether a target cell was treated with DMSO or not. (A) Cells were pretreated with DMSO for 15 minutes and later introduced into the sensing system at t=0 . The ten points moving average of the shift was 6.43 kHz at the beginning, later, declined to 4.64 kHz after half an hour (B) Control run with the same device. This time cells were treated with the culture medium instead of DMSO for 15 minutes and the change in the signal average remained close to initial levels (5.46

kHz to 5.11 kHz). (C) Optical image of a cell after DMSO treatment. (D) Optical image of a cell used in the control run. (E) Median size of cells with DMSO treatment and non-treatment.

Table III summarizes some of the literature on microwave sensing applied to cell measurements. Since different physical parameters are used in different studies, In surveying different sensors, one challenge is that different physical parameters (capacitance change, changes in S21 amplitudes, frequency shifts etc.) are used in different studies. Even within the same measurement modality, differences between sensor characteristics makes direct comparisons difficult. For instance, minimum resolvable capacitance can be used as a metric; however, the amount of capacitive change a particle induces depends on sensor architecture (e.g. the enhancement factor of the local Electrical field at the particle location could be different). For this reason, the resolution of single-particle measurements can serve as a more broadly applicable figure-of-merit for sensors. Yet in practical applications, single-particle resolution may not eventually recapitulate the performance levels, since other sources of uncertainty —such as the spread in the vertical position of analytes— can produce additional spread in the measured signals. For this reason, the ensemble resolution seems to be the final test bed in comparing different sensing platforms.

## 5. **Conclusion**

In conclusion, we showed several proof-of-concept demonstrations of microwave sensors in different sensing scenarios in microfluidic systems. When microwave sensors are used in conjunction with optical microscopy, both the geometric size of a particle and electrical capacitance change induced by the same particle can be obtained simultaneously. With the two-dimensional analytic information, it becomes possible to distinguish particles of different material compositions based on their intrinsic dielectric properties.

With CPW resonators, single-particle resolution levels (induces frequency shift over frequency noise) close to 100 were obtained for 20 μm PS particles in both sensor types. However, the ensemble resolution levels were much lower (10-18) due to the difference in height as particles passed through the sensing region. The work also demonstrates that standalone microwave sensors can be used to track the internal changes even before they are manifest visually. These results indicate that microwave sensors can add a unique analytical dimension for material screening and single-cell biophysics experiments

TABLE III

SUMMARY OF VARIOUS MICROWAVE BASED CELL SENSING EXPERIMENTS

| Reference | Measured Parameter | Measured Sample | Resolution Level / Demonstration |
| --- | --- | --- | --- |
| [28] ,[29] | Capacitance Change | 5.7 um PS microparticles<br>Yeast Cells, Chinese hamster ovary cells | 0.65 aF, 3 ms response time |
| [30] | S21 magnitude and resonance frequency at the same time | 5.8 um yeast cells | Live vs dead cells in a 2D parameter space of S21 and frequency shift. |
| [31] | S11 magnitude and resonance frequency at the same time | Cluster of several cells<br>(Stem Cells; U87 Glial Cells) | 1 MHz resolution where the signal is 100 MHz shift for 14.4 GHz (and 0.5 dB decrease in S11.) |
| [32] | Frequency shift at 4 different microwave frequencies | Different mammalian cell lines: MCF-10A, MDA-MB-231, THP- 1, and K-56, and polystyrene particles 10 to 0.5 um . | 1 aF capacitance change<br>5 ppm frequency change at 100 kHz signal bandwidth. Intrinsic CV (app. 10%) of the sample is attained. |
| [33] | S21 | Cell: C2C12, HMSC, Monocytes<br>Particle: 11um PS, 25um PS | C2C12: -0.0015 dB ; HMSC: -0.0011 dB; Monocytes: -0.0007<br>25 um PS: -0.013 dB; 11 um PS: -0.0008 dB |
| [34] | Insertion Loss Change (dB) at wide bandwidth. | Jurkat Cells | Fitted cell capacitance of 1.5 pF; uncertainty level of 0.3 pF |
| [35] | Capacitance Change | B lymphoma cell | 10 aF resolution for cells inducing 530 aF signal. |
| [36] | Capacitance Change | THP1 cells | 0.6 fF signal level for for living cells |
| [37] | Complex permittivity change<br>(dielectric constant, DK, & dielectric loss, DL) | Uveal Melanoma Cancer Cell (92.1 line) | DK change: 2.7<br>DL change: ~1 (with respect to initial DK and DL) |
| [56] | Frequency Shift<br>Microstripline Resonator | Mammalian cell lines HeLa | Allan Deviation 2 $10^{-8}$<br>Single cell resolution ~ 10 |
| This work | Frequency Shift<br>CPW and SRR resonators | Mammalian cell lines and 10-20 um polystyrene | Allan Deviation 4 $10^{-8}$ for CPW, $7.510^{-7}$ for SRR.<br>Single particle SNR ~ 100 for 20 um polystyrene |

ACKNOWLEDGMENT

This project has received funding from the European Research Council (ERC) under the European Union's Horizon 2020 research and innovation programme (grant agreement No 758769). We thank Ilbey Karakurt, Mehmet Kelleci and Hande Aydogmus for useful discussions.

NOTES

‡ Current Adress: Department of Applied Physics, Caltech, Pasadena, 91125, CA

§ Current Address: Medical University of South Carolina, Charleston, 29425, SC

* Corresponding Author: selimhanay@bilkent.edu.tr

## METHODS

**Microwave Measurement.** The resonance frequency of the chip was determined considering the S21 response of the sensing region with circulator (PE8401) on the vector network analyzer (Rohde & Schwarz ZNB40). The driving signal (around 2 - 3 GHz) from the signal generator (Rohde & Schwarz SMB 100A) was first up-converted with the output signal (2 - 3 MHz) from the lock-in amplifier (Zurich Inst. MFLI). Then, the signal was sent to the circulator to effectively drive and read the response of the resonator. Similarly, the response was down-converted and filtered with a low pass filter (MiniCircuits 4.5 MHz)

**Optical Size Measurements.** To obtain the optical size of each cell under bright field microscopy, the number of pixels inside each single cell was estimated digitally using *polyarea* function of MATLAB. The pixel size was matched to the dimensions of the device on the same image. The mean and standard deviation of the size of each cell/particle was obtained by measuring the same cell on many successive frames. The pixelated area was then converted first into the geometric area, and then assuming a spherical shape, into the volume of the particle/cell. For fluorescent cell analysis, image processing algorithm is explained in Supplementary Note 3.

**Cell Culturing.** MDA-MB-231 was purchased from ATCC (Manassas, VA, USA) and was labeled with green fluorescent protein (GFP) to be able to visualize under fluorescent light. Cells were cultured in Dulbecco's modified

Eagle's medium (Biowest, Nuaille, France), supplemented with 10% fetal bovine serum (FBS, Biowest), 1% non-essential amino acids (Biowest) and 50 U/ml penicillin/streptomycin (P/S, Biowest). The presence of mycoplasma in cells was tested regularly using MycoAlert Mycoplasma Detection Kit (Lonza, NJ, USA). The cells were incubated at 37 °C incubator (Thermo Scientific, IL, USA) under a humidified atmosphere of 5% CO2. For the standard experiments, cells were washed with phosphate-buffered saline (PBS, Biowest) and trypsinized (Biowest) to detach from the plates. The suspended cells were added to a falcon tube and centrifuged. The supernatant that contains trypsin was removed and fresh cell medium is added. cells were then transported to the sensing region with a controllable pressure pump (Fluigent; MFCS-EZ) and passed through a flow sensor with typical flow rates of 0.3 - 1 μL/min. To assess the changes in the size of the cells, cells were pre-treated with 100% pure DMSO (ChemCruz, TX, USA) and growth medium (DMEM) as a control group for 15 minutes and then collected for the analysis.

**Polystyrene Samples:** We used 10 ± 0.15 μm polystyrene particles (Sigma-Aldrich, Supelco - 72822), 20 $\pm$ 0.3 μm polystyrene particles (Sigma-Aldrich, Supelco - 74491) which constitutes a more homogeneous sample compared to a cell culture. We diluted the original solution with deionized water at a ratio of 100:1. A cell solution was suspended following the same steps in the previous experiments of Figure 2. Both microparticle and cell experiments were performed using the same device.

**SNR Calculations.** In experiments with 20 μm polystyrene particles (Sigma-Aldrich, Supelco - 74491), for each channel height, the maximum frequency shifts induced by each particle ($f_i$) is measured. The mean frequency shift is calculated as $\bar{f} = \frac{1}{N}\sum_{i=1}^{N} f_i$ where *N* denotes the number of microparticles measured for a given channel height (for all cases *N*>50). The Ensemble Standard Deviation ($\sigma$) is calculated as $\sigma = \sqrt{\frac{1}{N}\sum_{i=1}^{N}\left(f_i - \bar{f}\right)^2}$. Noise Level is obtained by calculating $\sqrt{2} \times \sigma_A(\tau_{FLL})$ where $\sigma_A(\tau_{FLL})$ stands for the Allan Deviation at the FLL time scale $\tau_{FLL}$, just prior to introducing microparticles into the sensor. SNR in frequency shift is calculated as the mean frequency shift over the noise level. SNR of the Ensemble is calculated by dividing the Mean Frequency Shift to Ensemble Standard Deviation. The dispersion of the commercial microparticle sample is reported as $\sigma_R = 0.3\ \mu m$ in diameter, with a mean diameter of $\bar{R} = 20.1\ \mu m$. The microwave frequency shift signals are proportional to the mass (and not the radius) of the microparticles. The mass of the microparticles scales as the cube of the diameter ($m \sim R^3$), hence the standard deviation of the mass distribution ($\sigma_m$) is $\sigma_m = 3\bar{m} \times \frac{\sigma_R}{\bar{R}}$. Here, $\bar{m}$ denotes the mean value for the mass, and the relationship is obtained by using standard formulas for relating the variances of random variables with narrow distributions. Plugging in the numbers, $\frac{\sigma_m}{\bar{m}} = 3 \times \frac{0.30\ \mu m}{20.1\ \mu m} = 0.045$. As a result, the SNR that can be obtained from this sample cannot be larger than $\frac{\bar{m}}{\sigma_m} = 22.3$ . These values are reported on Table I.